# Time-Series Analysis of Photovoltaic Distributed Generation Impacts on a Local Distributed Network[*]

Mir Hadi Athari, Zhifang Wang[†], Seyed Hamid Eylas
Department of Electrical and Computer Engineering
Virginia Commonwealth University
Richmond, VA, USA
Email: {atharih},{zfwang},{eylassh}@vcu.edu

*Abstract*—Increasing penetration level of photovoltaic (PV) distributed generation (DG) into distribution networks will have many impacts on nominal circuit operating conditions including voltage quality and reverse power flow issues. In U.S. most studies on PVDG impacts on distribution networks are performed for west coast and central states. The objective of this paper is to study the impacts of PVDG integration on local distribution network based on real-world settings for network parameters and time-series analysis. PVDG penetration level is considered to find the hosting capacity of the network without having major issues in terms of voltage quality and reverse power flow. Time-series analyses show that distributed installation of PVDGs on commercial buses has the maximum network energy loss reduction and larger penetration ratios for them. Additionally, the penetration ratio thresholds for which there will be no power quality and reverse power flow issues and optimal allocation of PVDG and penetration levels are identified for different installation scenarios.

## I. INTRODUCTION

Vast transmission and distribution networks increase power loss and make power systems less efficient. Environmental concerns and less reliance on fossil fuel power generation have caused a transition from centralized energy infrastructures to integration of distributed energy resources (DER) especially solar distributed generation. With recent advances in technology, solar PV is currently the fastest growing renewable energy source in the U.S. [1] becoming more accessible, affordable, and prevalent in the country than ever before. More than one million solar installations are connected to US electric grid today, totaling nearly 30 gigawatts – a 26-fold increase since 2008. The President's Climate Action Plan set the goal of doubling U.S. renewable energy deployment between 2012 and 2020 and installing 100 megawatts of solar on federally-assisted housing [2].

Extensive integration of PV distributed generation (PVDG) into distribution networks can result into disruption of the normal operation. Considering the radial topology of distribution networks and the fact that they are designed for centralized generation and unidirectional power flow, increasing penetration level of PVDG into these networks may arise numerous issues that need to be addressed [3], [4]. One of these issues is reverse power flow, which can affect voltage regulators and protection devices [5], [6]. During a fault, protection devices may not be able to detect fault currents as the bidirectional power flow from PVDG may reduce fault current and the relay will not trip. A loss of coordination between protective devices may also result [7]. These voltage fluctuations and rises result in increased voltage regulator and tap operations, which significantly decrease their lifetime of use [8]. Another issue triggered by extensive penetration of PVDG into distribution network is voltage quality problems. Although sunlight presents an abundant source of energy, its intermittency from cloud coverage can result in a high degree of voltage fluctuation, especially given that there is not rotating inertia present as in wind turbines. As a result, power output can change by 80% in just a few seconds [9]. Such rapid voltage fluctuations can lead to load imbalances, voltage flickers, and undervoltage [10]. Also, overvoltage due to high PV output power during solar peak irradiance hour may occur which can cause damage to sensitive loads.

During recent years a tremendous effort has been put on studying the impacts of PVDG on distribution network from both utility companies and scholars. Couple studies investigated these impacts for local networks particularly for U.S. and presented various remedial actions. Many of the studies cover California as the lead state in number of solar projects installed over past decade [11]–[14]. The PV integration limit, termed hosting capacity, is calculated with respect to bus overvoltages, voltage deviations, and voltage unbalance using California solar statistics [11]. Another study has utilized real field data of a sample circuit from Southern California to emphasize the necessity of time-series simulation in understanding true impacts of high penetration PVDG installation into distribution networks [12]. Cohen and Callaway studied the effects of PVDG on California's distribution system both from engineering and economic points of view [13], [14]. Others have designed feeder models based on real world designs. Baran *et al.* examined a distribution

---

system in Raleigh, NC to investigate how PV affects protection devices [5]. Hill *et al.* based their research on a battery energy storage scheme on a Hardware-in-the-loop test bed in Texas [15].

However, there is still a need for studies on PVDG impacts on large-scale real-world feeders that incorporates real time solar insolation data along with time-series analysis in other local areas in U.S. Our paper aims to investigate PVDG impacts on a local distribution network by incorporating feeder designs from Dominion Virginia Power and real world solar irradiation time-series data and different load profiles including residential, commercial, and industrial loads to provide accurate models of the impact PVDG has on a local distribution network. In addition, we identify PVDG penetration ratios for which the reverse power flow and overvoltage problems come into picture and then we provide recommendation for optimal allocation and maximum penetration levels for PVDGs in order to avoid such problems.

The rest of the paper is organized as follows. In section II the description of multi-feeder model is presented. Section III describes the PV panel model. Simulation scenarios and optimal allocation and penetration levels are discussed in section IV and section V presents several mitigation strategies and future frameworks and concludes the paper.

## II. DESCRIPTION OF MULTI-FEEDER MODEL

IEEE 69 bus distribution network system is selected to perform simulation for distributed PV. The schematic diagram of the networks is shown in Fig.1. This system is consisted of one main feeder (F1) plus 7 additional braches (F2-F8) supplying three different types of load including residential, commercial, and industrial loads. The network voltage level is 12.66 kV.

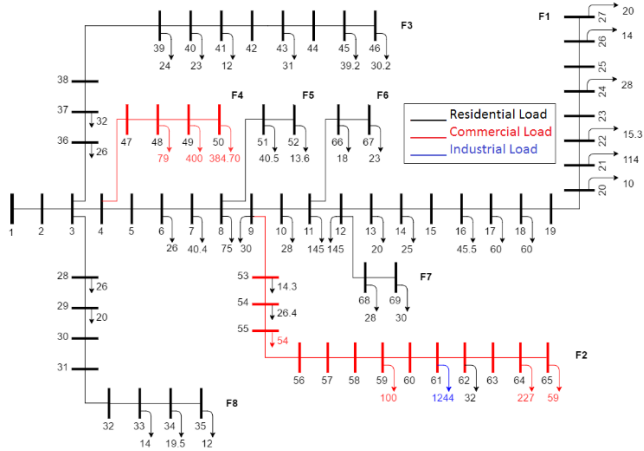

Figure 1. Schematic diagram of redesigned 69 bus distribution network.

The average load of each bus is shown in Fig. 1 with specified color code and the number representing active power in kW. The total average load is $S_L = 5.4 + 3.8\ MVA$ which is consisting of three types of loads including residential, commercial, and industrial loads. The distribution system is redesigned and adapted to accommodate for inputs from Dominion Virginia Power staff in terms of voltage settings and feeder length-load settings as well as residential/ commercial/ industrial demands profiles. Note that the allocation of load types for each feeder and their placement is selected based on their sizes according to comments from Dominion experts. Each feeder supplies various types of loads (e.g. feeder 2) or only residential demands (e.g. feeder 1). The physical and electrical parameters of the network is determined based on power flow results considering allowable range for voltage profile and current flowing through each line. Finally, depending on each type of load (residential/ commercial/ industrial), load profiles are generated from hourly data available at [16] for Richmond, Virginia. Fig. 2 shows the load profiles for three types of loads considered for this study.

After a statistical evaluation of power factor for residential and commercial loads from original setting of IEEE 69 system, it is found that the power factor for residential loads follows a normal distribution with μ=0.8158 and $\sigma$= 0.0181 and for commercial loads it represents a uniform distribution in the range of [0.8110, 0.8150]. Next, given the active demanded power for each load, the reactive power profile is calculated according to random power factors generated from normal and uniform distributions. Next, the goal is to study the impact of PVDG on voltage and current profiles as well as network real power loss. Different scenarios are considered to investigate the impacts of different penetration ratio and configuration of distributed PV.

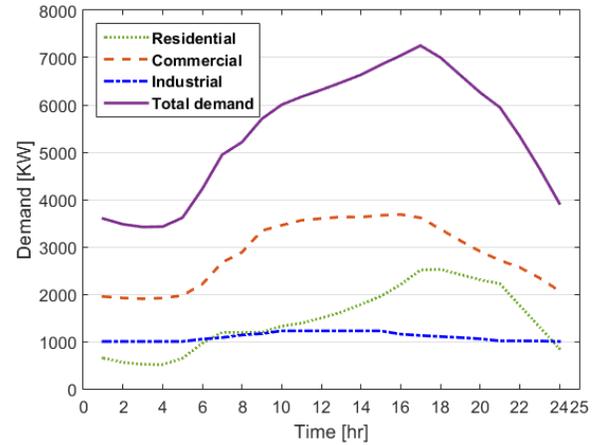

Figure 2. Load profiles utilized in time-series simulation for different load types.

## III. PV PANEL MODEL

With consideration for the effect of temperature, the output generated power (kW) by PV array with $N_s$ modules in series and $N_p$ modules in parallel is given by [17]

$$P_{pv}(t) = N_s . N_p . \frac{\frac{V_{OC}}{n_{MPP}KT/q} - \ln\left(\frac{V_{OC}}{n_{MPP}KT/q} + 0.72\right)}{1 + \frac{V_{OC}}{n_{MPP}KT/q}} . \left(1 - \frac{R_s}{V_{OC}/I_{SC}}\right) . I_{SC0} \left(\frac{G}{G_0}\right)^\alpha . \frac{V_{OC0}}{1 + \beta \ln\frac{G_0}{G}} . \left(\frac{T_0}{T}\right)^\gamma . \eta_{MPPT} \eta_{oth} \quad (1)$$

where $n_{MPP}$ is the ideality factor at the maximum power point ($1 < n_{MPP} < 2$), $T$ is temperature of the PV module (K), $K$ is Boltzmann constant ($1.38 \times 10^{-23}$ J/K), $q$ is the magnitude of electron charge ($1.6 \times 10^{-19}$ C), $R_s$ is the series resistance (Ω), $\alpha$ is the exponent responsible for all the non-linear effects that the photocurrent depends on, $\beta$ is a PV module technology specific-related dimensionless coefficient, and $\gamma$ is the exponent considering all the non-linear temperature–voltage effects. $V_{OC}$ & $I_{SC}$ are open-circuit voltage (V) and short-circuit current (A) of the PV module which has been calculated under different solar irradiance intensities ($G$, $G_0$) and two PV module temperatures ($T$, $T_0$), respectively. Finally, $\eta_{MPPT}$ is maximum power point tracker efficiency and $\eta_{oth}$ is the factor representing other losses such as the loss caused by cable resistance and accumulative dust. The characteristic parameters of PV modules used in this study is presented in Table I [17].

The PV module can be placed at any orientation and at any slope angle, but most local observatories only provide solar radiation data on a horizontal plane. Thus, an estimate of the total solar radiation incident on the PV module surface is needed. Generally, the total solar radiation on a tilted surface is calculated by adding the beam, diffuse and reflected solar radiation components on the tilted surface:

$$G_{total} = G_{bt} + G_{dt} + G_{re} \quad (2)$$

where $G_{total}$ is the total solar radiation on a tilt surface; $G_{bt}$, $G_{dt}$ and $G_{re}$ are the beam, diffuse and reflected radiation on the tilt surface. The irradiance data for this study is collected from [18] and are shown in Fig. 3 for a four different PV sites used in simulation process.

TABLE I. CHARACTERISTIC PARAMETERS OF PV MODULE

| Parameter | Value | Parameter | Value |
|---|---|---|---|
| $\alpha$ | 1.21 | $V_{oc}$ (V) | 21 |
| $\beta$ | 0.058 | $G_0$ (W/$m^2$) | 1000 |
| $\gamma$ | 1.15 | $T_0$ (K) | 298 |
| $n_{MPP}$ | 1.17 | $\eta_{MPPT}$ | 0.95 |
| $R_s$ (Ω) | 0.012 | $N_{Cell,El}$ | 24 |
| $I_{sc}$ (A) | 6.5 | - | - |

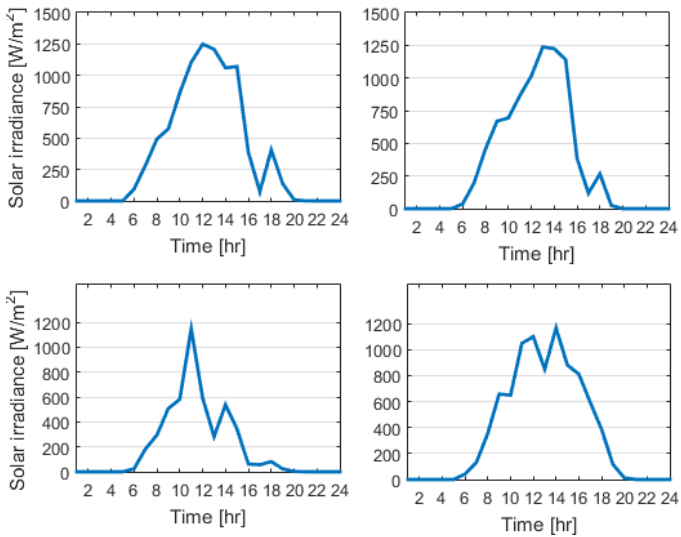

Figure 3. Solar irradiance data for a select PV site.

IV. SIMULATION SCENARIOS AND RESULTS

Three scenarios based on PV installation configuration and penetration ratio of PVDG are considered to study the impacts of PVDG on voltage and current profiles and network power loss. For the first scenario impacts of utility scale centralized installation of PVDG are investigated considering different penetration ratio to identify hosting capacity of the network. Then, two more scenarios are considered to study the impacts of high penetration PVDG on residential and commercial/industrial consumers. These scenarios allow us to see the impact of more distributed installation of PVDG on voltage and current profiles and network power loss. For later cases, the installation capacity for PVDG is proportional to the load on each bus to take into account the physical constraints of PVDG integration. This means that larger loads (households or commercial units) will have larger PV installation. Note that simulations for all three scenarios involve hourly data for load profiles and solar irradiance, thus, the fast dynamics such as cloud impacts and voltage flickering cannot be examined.

In this paper PV penetration ratio is defined based on sub-station peak load and is as follows

$$PR(\%) = \frac{\sum_{i=1}^{N} P_{PV_i}^{max}}{\max(\sum_{j=1}^{M} P_{Load_j})} \quad (3)$$

where $P_{PV}$ and $P_{Load}$ are PV panel output power (kW) and electrical load demand (kW), respectively. Fig. 4 shows the voltage and current profiles of feeder 1 and 2 without any PV integration into grid with time-space representation. As shown in the network diagram in Fig. 1, the main feeder (F1) supplies mainly residential loads where we can see the pattern of typical residential load with larger peak around 8:00 PM in its current profile. While feeder 2 mainly supplies large commercial and industrial loads with a different pattern than residential loads. It's found that commercial and industrial loads tend to peak around 1:00 PM which is concurrent with peak solar irradiance for this location.

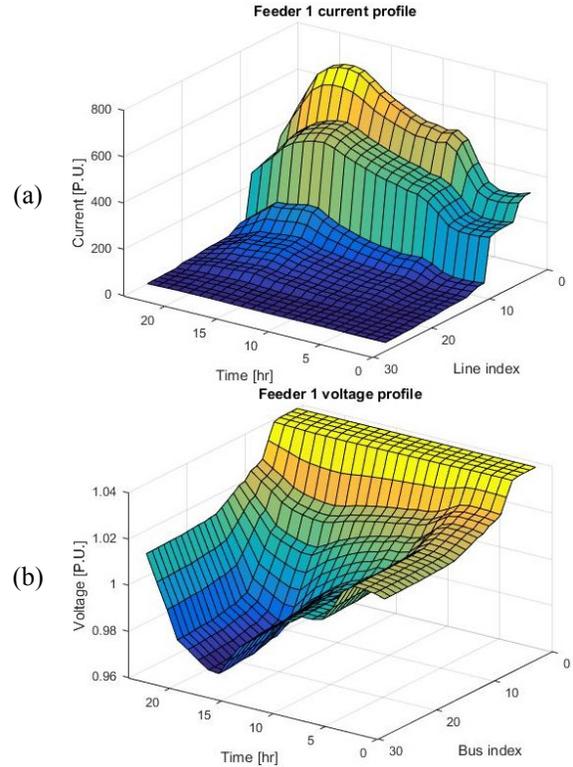

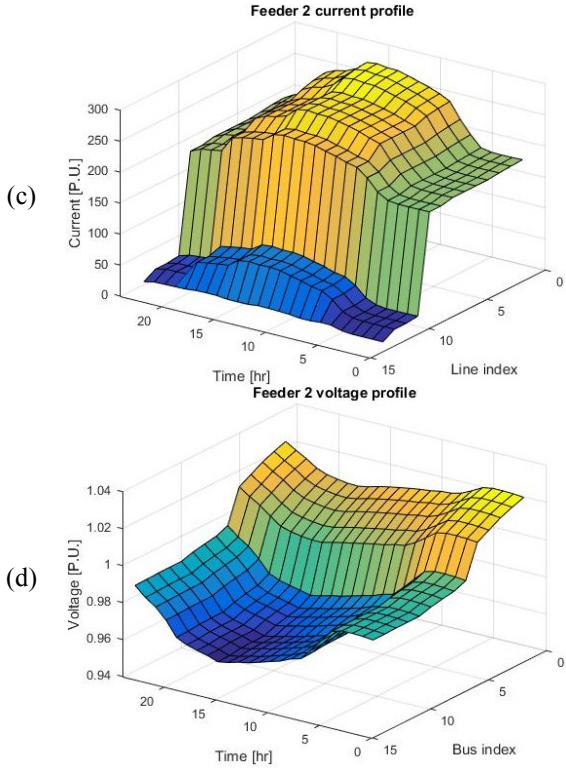

Figure 4. Voltage and current profiles for feeder 1 (a,b) and feeder 2 (c,d) without PV penetartion.

### A. Utility Scale Centralized PVDG installation

Three different configurations are considered to study the impacts of utility scale centralized PVDG installation as shown in Table II. For the first configuration, three utility scale PVDGs are installed at the beginning of the feeder on buses 4,5, and 6 allowing us to increase penetration level up to 75% without facing reverse power flow issues. The total energy loss of the network for different penetration levels for all scenarios are presented in Table II. The total energy loss of the network before installing any PVDG was 4845 KWh. There is a maximum energy loss reduction of 5% for the first configuration without having any voltage quality issue and reverse power flow. For this installation scenario, output power of PV sites during peak solar irradiance contributes to the main feeder and decreases the magnitude of current flowing in lines near the beginning of the feeder as shown in Fig. 5. Thus, there is no noticeable reduction in network energy loss however, with this installation the utility can increase installed capacity of PVDG up to 75% without having voltage quality or reverse power flow issues. However, the loss reduction for this configuration is smaller compared to the other two configurations where PVDGs are installed at the middle and end of the feeder. Note that, for latter cases, the loss reduction is bigger for low penetration ratios while going beyond a certain ratio (30%) results into increased network loss and also reverse power flow and overvoltage issues.

### B. Distributed Commercial Installation

For this scenario PVDGs are installed on all 7 commercial buses mainly located in feeder 2. Since the load profile of commercial loads better match the PV output power, there is considerable total energy loss reduction for this scenario with a maximum 40% reduction with 56% penetration ratio without having any voltage quality or reverse power flow issues. Fig. 6 shows the voltage and current profiles for feeder 2 with penetration ratio of 15%. We can see a better peak shaving performance for PVDG in this scenario as the peak load and peak PV output are almost concurrent. We can also see a boost in voltage level along the feeder around noon which delivers a smoother voltage profile and improves voltage quality for this feeder.

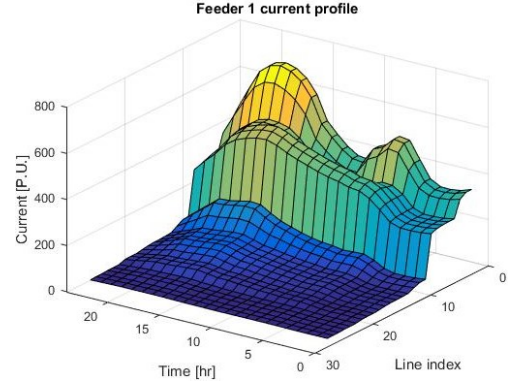

Figure 5. Feeder 1 current profile for centralized scenario with 50% penetration ratio.

### C. Distributed Residential Installation

For this scenario the highly distributed installation for PVDGs is investigated. All residential buses have PV sites installed so that the installation capacity is proportional to load size. In this way the physical limitation on rooftop area for households is taken into account.

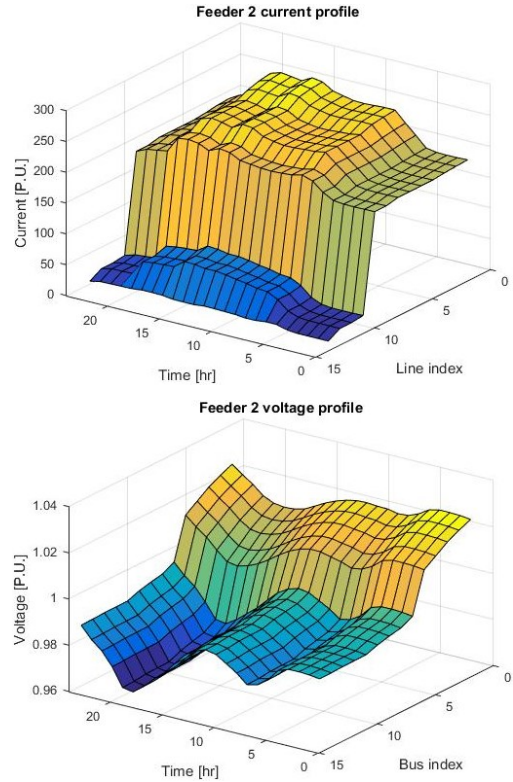

Figure 6. Feeder 2 voltage and current profiles with 15% PVDG penetration ratio for commercial installation scenario.

TABLE II. TOTAL ENERGY LOSS FOR INVESTIGATED SCENARIOS WITH DIFFERENT PENETARTION RATIO

| Scenarios | Penetration ratio | 5% | 10% | 20% | 30% | 50% | 100% | PR*(%) | PR+(%) |
|---|---|---|---|---|---|---|---|---|---|
| | Installation location (bus number) | Total energy loss (kWh) | | | | | | | |
| Utility scale centralized installation I | 4,5,6 | 4836 | 4829 | 4796 | 4771 | 4750 | 4670 | 75% | 80% |
| Utility scale centralized installation II | 15,16,17 | 4769 | 4685 | 4520 | 4434 | 4686 | 6368 | 11% | 35% |
| Utility scale centralized installation III | 24,25,26 | 4733 | 4588 | 4494 | 4551 | 4850 | 7977 | 5% | 25% |
| Distributed commercial installation | 48-50,54,59,64,65 | 4582 | 4237 | 4039 | 3762 | 3070 | 2816 | 56% | 93% |
| Distributed residential installation | 6-46,51,52,53,54,62,66,67 | 4760 | 4674 | 4519 | 4387 | 4205 | 4153 | 25% | 75% |

\* Penetration ratio with first observed reverse power flow
+ Penetration ratio with first observed overvoltage

As shown in Table II total loss reduction for this scenario is larger than centralized installation at the begging of the feeder however it is smaller compared to loss reduction achieved with distributed commercial installation. Fig. 7 shows the voltage and current profiles for feeder 1 with 30% penetration ratio of PVDG. As indicated in the figure, reverse power flow occurs for this penetration ratio for lines 10-20 in the main feeder. This reveres flow can interrupt the operation of protective devices by decreasing the fault current magnitude thus leading to undetected faults in the distribution network.

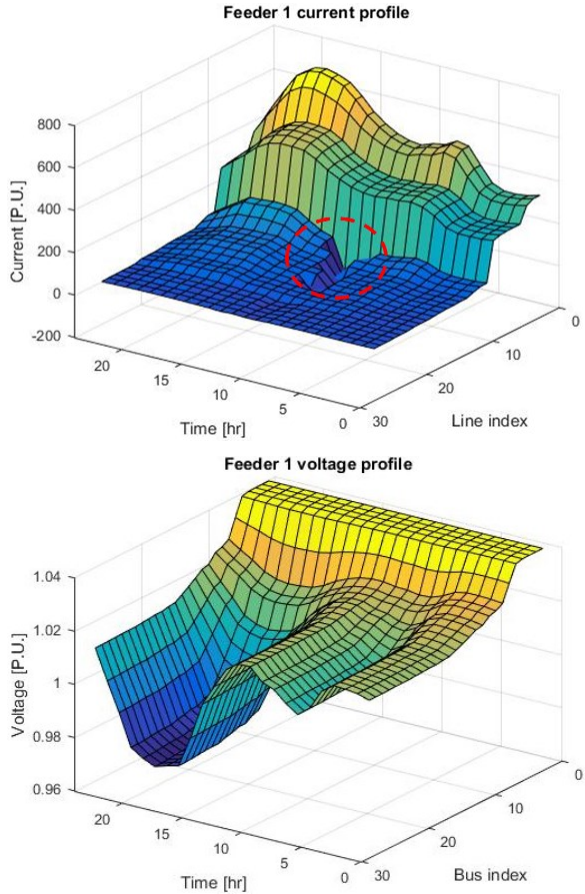

Figure 7. Feeder 1 voltage and current profiles with 30% PVDG penetration ratio for residential installation scenario.

### D. Optimal placement of PV sites

Results presented in Table II suggests that for different scenarios there are maximum penetration level for which there will be no reverse power flow and overvoltage issues. Also, for centralized utility scale configurations, the loss reduction and penetration levels highly depend on installation locations. Hence, it is necessary to solve an optimization problem to find the best penetration ratio for each scenario without having the two issues. For distributed installations, we detected the penetration levels for which the reverse power flow and overvoltage problems begin as shown in Table II. For both scenarios the penetration level associated with first reverse power flow in the network is smaller than the penetration level for first overvoltage in the network due to high penetration of PV. Thus, the former is set as the maximum allowable penetration ratio for each scenario to avoid both reverse power flow and overvoltage problems.

To find the optimal settings for PV sites for each scenario, an optimization using particle swarm optimization (PSO) algorithm is performed to minimize the total energy loss of the network and voltage deviation. The optimal settings for each PV site include its placement and installed capacity. For each installation configuration, the total real energy loss of the network will be calculated according to [19]:

$$E_{loss} = \sum_{t=1}^{24}\sum_{l=1}^{L} |i_L^t|^2 R_L \quad (4)$$

where $i_L^t$ is current flowing through line $L$ at time $t$ and $R_L$ is resistance of line $L$. The formulation of voltage profile deviation from nominal value with $v_n^t$ as the voltage of bus $n$ at time $t$ is as follow [20]:

$$v_D = \sum_{t=1}^{24}\sum_{l=1}^{n} (v_i^t - 1)^2 \quad (5)$$

Finally, the multi-objective function is formulated as

$$\min_{L_{PV}, P_{PV}^{max}} f = E_{loss} + \omega * v_D \quad (6)$$

Subject to:

$$i_{ij}^t \geq 0 \quad \forall\, i < j \quad (7)$$

$$0.95 \leq |v_i| \leq 1.05 \quad (8)$$

where $\mathcal{L}_{PV} = [\ell_1, \ell_2, ..., \ell_n]^T$ $\ell_i \in \{0,1\}$ is PVDG location vector, $\omega$ is weighting factor, and $P_{PV}^{max} = [P_1^{PV}, P_2^{PV}, ..., P_n^{PV}]$ is PVDG maximum capacity vector. The minimum of the above objective function ensures the optimal settings for PVDGs with minimum power loss and enhanced voltage profiles for the network.

Table III shows the total energy loss and voltage deviation ($v_D$) associated with the optimal settings for each scenario. Note that optimal sizing and sitting are considered for the maximum penetration level for distributed installations without reverse power flow issues. It is found that buses 17, 18, and 19 are the best locations for centralized installation with maximum network loss reduction and minimum voltage deviation. For distributed installations it is found that by optimal setting of PVDGs we can achieve even more energy loss reduction and also have enhancement in voltage profiles compared to PVDG size allocation based only on bus load size.

TABLE III. OPTIMAL ALLOCATION AND PENETRATION LEVEL RESULTS FOR STUDIED SCENARIOS

| Installation Scenarios | Optimal PR | Total energy loss | Voltage deviation ($v_D$) |
|---|---|---|---|
| Utility scale centralized installation | 10% | 4545 | 1.7359 |
| Distributed commercial installation | 55% | 3013 | 1.6954 |
| Distributed residential installation | 25% | 4452 | 1.6123 |

V. CONCLUSION AND FUTURE WORKS

The impacts of photovoltaic distributed generation (PVDG) on a local real-world distribution network in Richmond, Virginia area are studied. The parameters investigated are voltage and current profiles for each feeder and total network energy loss. Different scenarios based on installation location and penetration ratio of PVDGs are considered to cover more common scenarios most utilities are interested in. Actual load profiles for different load types including residential, commercial, and industrial consumers along with historical solar irradiance data for the area are considered to perform time-series analysis and to evaluate the impacts of solar peak power on network voltage and current profiles. Finally, the optimal allocation of PVDGs and penetration levels without reverse power flow and overvoltage problems are identified for each scenario.

It is found that the maximum energy loss reduction can be achieved with distributed installation for commercial buses because of better alignment of commercial loads profile with solar irradiance during the day. However, for penetration ratios above a certain threshold there would be voltage quality and reverse power flow issues that need to be addressed. Hence, optimal sitting of PVDGs suggests a maximum penetration level to avoid such problems and having the maximum network power loss reduction and minimum voltage deviation. For further mitigation approaches we can consider demand management of customer loads, fast voltage regulation/control via distributed VAR compensation, and energy storage.

As the future extension of this study it is of interest to incorporate the data with finer resolution and transient models of distribution network and electronic devices to examine the fast dynamics such as cloud impacts and voltage flickering.